\documentclass[12pt]{iopart}

\usepackage{graphicx}
\usepackage{epsfig}

\begin{document}

\title{Phase space picture of Morse-like coherent states based upon the Wigner function}
\author{O de los Santos-S\'anchez$^{(1)*}$ and J R\'ecamier$^{(1)}$} 

\address{$^{(1)}$ Instituto de Ciencias F\'{\i}sicas, Universidad
  Nacional Aut\'onoma de M\'exico, Apdo. Postal 48-3, 
Cuernavaca, Morelos 62251, M\'exico}

\eads{\mailto{octavio.desantos@gmail.com}}

\begin{abstract}
Using the Wigner distribution function, we analyze the behavior on phase space of generalized coherent states associated with the Morse potential (Morse-like coherent states). Within the f-deformed oscillator formalism, such states are constructed by means of the two following definitions: {\it i)} as deformed displacement operator coherent states (DOCSs) and {\it ii)} as deformed photon-added coherent states (DPACSs). 
\end{abstract}

\maketitle

\section{Introduction}

The notion of coherent states is an important concept that comes from the study of the quantum harmonic oscillator. It is a well-known fact that coherent states are the quantum states which most closely resemble classical states. They were first introduced by Schr\"odinger \cite{mandel} who referred to them as states of the minimum uncertainty product, and are such that when being subjected to a simple harmonic potential, they remain well localized around the corresponding classical trajectory and always preserve the minimum uncertainty product for the position and momentum canonical variables. More than three decades after Schr\"odinger's idea, Glauber \cite{glauber} showed that coherent states are also useful for describing the quantum properties of the electromagnetic field. 
Since then, such states have played a preponderant role in quantum optics both theoretically and experimentally \cite{yamamoto}. \\

According to Glauber \cite{glauber}, the so-called field coherent states can be constructed from any one of the following definitions: {\it i)} as eigenstates of the annihilation operator operator $\hat{a}$ of the harmonic oscillator, {\it ii)} as those states obtained by the application of the displacement operator $\hat{D}(\alpha)=\exp(\alpha \hat{a}^{\dagger}-\alpha^{\ast}\hat{a})$ upon the vacuum state, and {\it iii)} as the quantum states that fulfill a minimum-uncertainty relationship. When one makes use of the harmonic oscillator algebra, the same coherent states are obtained from the three Glauber's mathematical definitions. However, for potentials other than the harmonic oscillator, each one of such definitions gives raise to different coherent states that may possess nonclassical properties. In the literature, there has been a great deal of interest in generalizing the concept of coherent states. A large number of proposals have arisen from the dynamical point of view,  by considering the underlying symmetry associated to the system of interest, and those based on deformed or nonlinear algebras.

Among the most representative generalizations, based on dynamics, a class of coherent states was constructed by Nieto and Simmons \cite{nieto1,nieto2,nieto3} for a number of systems, applying a methodology introduced by themselves in order to look for states that would imitate the classical motion in a given one-dimensional potential. They defined them as those states that saturate a generalized uncertainty relation. Gazeau and Klauder \cite{gazeau} proposed another class of coherent states for systems having a discrete and/or continuous spectrum. Such states possess continuity of labeling, a resolution of unity, and temporal stability.

As regards symmetry considerations, generalized coherent states for different Lie groups have been introduced. For instance, coherent states of the SU(1,1) Lie group were constructed by Barut and Girardelo \cite{barut} as eigenstates of the ladder operators. Puri and Agarwal \cite{puriagarwal} also constructed coherent states of SU(1,1) defining them as minimum uncertainty states, i.e., the so-called SU(1,1) intelligent coherent states. Independently, Gilmore \cite{gilmore1} and Perelomov \cite{perelomov} proposed a general algorithm to construct coherent states of dynamical groups for a given quantum system; such states are defined in terms of a displaced reference state.  

A novel algebraic generalization in terms of eigenstates of a deformed annihilation operator for f-deformed oscillators was effected by Man'ko \etal \cite{manko1,manko2}. It turns out that the coherent states thus obtained, also called nonlinear coherent states or f-deformed coherent states, display nonclassical properties like squeezing and anti-bunching effects; that is why they are usually referred to as nonclassical states. In fact, it is worth pointing out that this class of states naturally emerges from the description of the quantized center-of-mass motion of a laser-driven trapped ion (see \cite{matos} for more details). \\

In this work we are interested in studying the behavior on phase space of generalized coherent states associated with the Morse potential (Morse-like coherent states) considering just the bound part of the spectrum. To this end we will make use of the phase-space description based upon the Wigner distribution function \cite{kim}. It is known that, among many others properties, this mathematical tool provides a very good pictorial representation of the behavior of quantum states on phase space; for instance, either spreading wave packets or negativities in the distribution function are indicators of the nonclassical nature of the quantum system under consideration. To carry out the construction of our coherent states, we will make use of the f-deformed oscillator formalism developed by Man'ko and coworkers \cite{manko1,manko2}. One of the advantages of using this formalism is that one is able to choose a deformation function in such a way that the energy spectrum corresponding to our model Hamiltonian turns out to be essentially the same to that of a Morse potential. The details about this issue will be given below. \\

The photon added coherent states (PACS) introduced by Agarwal and Tara \cite{agarwal} are defined by the action ($m$ times) of the creation operator upon the harmonic-oscillator coherent state $|\alpha\rangle$. These states exhibit nonclassical features like squeezing and sub-Poissonian statistics. They were introduced as  intermediate states between the purely quantum Fock states and the coherent states.  In the limit $\alpha \rightarrow 0$ the state reduces to a Fock state, in the limit $m=0$ it reduces to a coherent state.
 If instead of using a coherent state $|\alpha\rangle$ one uses a nonlinear coherent state, for instance the one constructed as eigenstate of a deformed annihilation operator $\hat A = \hat a f(\hat n)$, then the PANLCS (Photon Added Non Linear Coherent States) would be defined by the application ($m$ times) of the deformed creation operator upon the Non linear coherent state \cite{safaeian}. \\

This paper is organized as follows. In section 2 we construct our Morse-like coherent states having in mind two kinds of generalization: (a) as those states obtained by the application of a deformed displacement operator upon the ground state of the oscillator (DOCSs) and (b) as  deformed photon-added coherent states (DPACSs) constructed by the iterative application of a deformed creation operator upon the states defined in (a). 
Analytic expressions of the Wigner distribution function for the aforementioned generalized coherent states are presented in section 3. In section 4 we plot some Wigner functions and discuss their temporal behavior on phase space. Finally, the conclusions are given in section 5.  

\section{Algebraic Hamiltonian for the Morse potential and its coherent states}

Firstly, let us introduce the one-dimensional Morse potential and their eigenfunctions. Such a potential, as a function of the position $x$ referred to the equilibrium point, is usually written in the form \cite{pot:morse}

\begin{equation}
V(x) = D \left[(1-e^{-\beta x})^{2}-1 \right],
\label{eq:morse}
\end{equation}
where its depth $D$ and its range $\beta^{-1}$ are related to the corresponding spectroscopic data of the molecule, i.e., 
\begin{equation}
D= \frac{\hbar \omega_{e}^{2}}{4\omega_{e}\chi_{e}}, \qquad  \omega_{e}\chi_{e} = \frac{\beta^{2}\hbar}{2 \mu },
\end{equation}
where, in turn, $\mu$ is the reduced mass of the molecule, and $\omega_{e}$ and $\omega_{e}\chi_{e}$ are, respectively, the harmonic and anharmonic spectroscopic parameters. 

It is known that  this potential possesses both a discrete and continuum spectrum, however the discrete part will be the one of interest to us. The corresponding energy spectrum is
\begin{equation}
E_{n} = \hbar \omega_{e} (n+1/2)-\hbar \omega_{e} \chi_{e} (n+1/2)^{2}.
\label{eq:spec1}
\end{equation}
The bound wavefunctions associated with the potential (\ref{eq:morse}) are given in terms of the associated Laguerre polynomials \cite{pot:morse}
\begin{equation}
\langle x|N,n\rangle = \psi_{N,n}(\xi(x)) = C_{N,n}e^{-\xi/2} \xi^{N-n} L_{n}^{2 N- 2n} (\xi),
\label{eq:wavefuncs}
\end{equation}
where $0 \le n \le N-1 $, with $N$ being the number of bound states, the normalization constant is 
\begin{equation}
C_{N,n} = \sqrt{\frac{2n! \beta (N-n)}{\Gamma(2N-n+1)}},
\end{equation}
and the Morse variable is defined as 
 \begin{equation}
\xi(x) = (2N+1)e^{-\beta x}.
\label{eq:mvariable}
\end{equation}

In accordance with the so-called f-deformed oscillator formalism introduced by Man'ko group \cite{manko1,manko2}, an f-deformed oscillator is a nonharmonic system characterized by a Hamiltonian of the harmonic oscillator form
\begin{equation}
\hat{H}_{f} = \frac{\hbar \Omega}{2}(\hat{A}^{\dagger}\hat{A}+\hat{A}\hat{A}^{\dagger}),
\label{eq:model1}
\end{equation}
where the deformed boson creation and annihilation operators $\hat{A}^{\dagger}$ and $\hat{A}$ are obtained by deforming the harmonic oscillator operators $\hat{a}$ and $\hat{a}^{\dagger}$  in such a way that 
\begin{equation}
\hat{A} = \hat{a}f(\hat{n}) = f(\hat{n}+1)\hat{a}, \qquad {\rm and} \qquad \hat{A}^{\dagger} = f(\hat{n})\hat{a}^{\dagger} = \hat{a}^{\dagger} f(\hat{n}+1),
\label{eq:defops}
\end{equation}
where $\hat{n}=\hat{a}^{\dagger}\hat{a}$ is the usual number operator and the operator function $f(\hat{n})$, which is assumed to be real (i.e., for real argument $x$ its value $f(x)$ is also real),  is a deformation function depending on the level of excitation.

The f-deformed oscillators can be construed as nonlinear oscillators whose nonlinearity is determined by the number-depending function $f$, while the operators $\hat{A}$ and $\hat{A}^{\dagger}$ are considered to represent the dynamical variables associated with this class of quantum systems \cite{mancini}. The deformed operators have the following effect on the number operator basis $|n\rangle$:
\begin{eqnarray}
\hat{A} |n\rangle & = & f(n)\sqrt{n} |n-1 \rangle, \\
\hat{A}^{\dagger} |n\rangle & = & f(n+1)\sqrt{n+1} |n+1\rangle. 
\end{eqnarray}
As we can see, these operators change the number of quanta in $\pm $1 and their matrix elements are modified through the deformation function $f(\hat{n})$. \\

It follows from the non-canonical transformation (8), together with the commutation relation $[\hat{a},\hat{a}^{\dagger}]=1$, that the deformed Hamiltonian can be rewritten in another convenient form as
\begin{equation}
\hat{H}_{f} = \frac{\hbar \Omega}{2}((\hat{n}+1)f^{2}(\hat{n}+1)+\hat{n} f^{2}(\hat{n})).
\label{eq:model2}
\end{equation}
In addition to this, it turns out that the set of operators $\{ \hat{A}^{\dagger},\hat{A}, \hat{n} \}$ obeys the commutation relations 
\begin{equation}
[\hat{A}^{\dagger},\hat{n}] = -\hat{A}^{\dagger}, \qquad [\hat{A},n] = \hat{A},
\end{equation}
and
\begin{equation}
[\hat{A}, \hat{A}^{\dagger}] = (\hat{n}+1)f^{2}(\hat{n}+1)-\hat{n}f^{2}(\hat{n}),
\end{equation}
where, depending on the explicit form of the deformation, the commutator of $\hat{A}$ and $\hat{A}^{\dagger}$ may become a rather complicated function of the number operator. \\

It can be easily seen that in the limit $f(\hat{n})=1$ one regains the harmonic oscillator algebra. On the other hand, and most importantly, by choosing conveniently the deformation function $f$, it may be possible to reproduce, with the help of the deformed Hamiltonian defined in (\ref{eq:model2}), the energy spectrum of the particular system under consideration. In the following we shall employ Eq. (\ref{eq:model2}) as an algebraic Hamiltonian in order to describe the Morse potential, and based on this algebraic formalism we shall construct our Morse-like coherent states.

\subsection{Deformed displacement operator coherent states}
It was shown in Ref. \cite{recamier1} that by choosing the deformation function 
\begin{equation}
f^{2}(\hat{n}) = 1-\chi_{a} \hat{n},
\end{equation}
with an anharmonicity parameter $\chi_{a} = 1/(2N+1)$, the deformed Hamiltonian (\ref{eq:model2}) turns out to be
\begin{equation}
\hat{H}_{D} = \hbar \Omega \left[\hat{n}+\frac{1}{2}-\chi_{a} \left(\hat{n}+\frac{1}{2} \right)^{2} -\frac{\chi_{a}}{4} \right],
\label{eq:spec2}
\end{equation}
whose spectrum is essentially the same, apart from an unimportant constant term, as that of a Morse oscillator (see Eq. (\ref{eq:spec1})). By comparing Eqs. (\ref{eq:spec1}) and (\ref{eq:spec2}) one can infer the relations $\omega_{e}=\Omega$ and $\chi_{e}=\chi_{a}$.

Given that we know the deformation function, the commutation relations for the deformed operators are explicitly given by
\begin{equation}
[\hat{A}, \hat{n}] = \hat{A}, \qquad [\hat{A}^{\dagger},\hat{n}] = -\hat{A}^{\dagger}, \qquad [\hat{A}, \hat{A}^{\dagger}] = 1-\chi_{a}(2\hat{n}+1).
\label{eq:crdef}
\end{equation}

For this particular choice of $f(\hat{n})$ the action of the deformed operators upon the energy eigenstates basis of the Hamiltonian (\ref{eq:spec2}), namely,
\begin{eqnarray} 
\hat{A} |n\rangle & = &  \sqrt{n(1-\chi_{a}n)} |n-1\rangle, \nonumber \\
\hat{A}^{\dagger} |n\rangle & = &  \sqrt{(n+1)(1-\chi_{a}(n+1))} |n+1\rangle, \label{eq:acdef}
\end{eqnarray}
is exactly the same as that of the actual lowering and raising operators upon the corresponding  Morse wave functions (\ref{eq:wavefuncs}) (see Ref. \cite{dong} for more details), i.e., 
\begin{eqnarray}
\hat{b} \psi_{N,n}(\xi) & = & \sqrt{n\left (1-\frac{1}{2N+1}n \right )} \psi_{N,n-1}(\xi), \nonumber \\
\hat{b}^{\dagger} \psi_{N,n}(\xi) & = & \sqrt{(n+1)\left(1-\frac{1}{2N+1}(n+1) \right)} \psi_{N,n+1}(\xi), \label{eq:ladder}
\end{eqnarray}
where the ladder operators $\hat{b}$, $\hat{b}^{\dagger}$ were obtained by means of traditional factorization methods, and whose differential form is given in terms of the Morse variable $\xi(x)$ as:
\begin{eqnarray}
\hat{b} & = & - \left [\frac{d}{d\xi} (2s+1)-\frac{1}{\xi} s(2s+1)+N+\frac{1}{2} \right] \sqrt{\frac{s+1}{s(2N+1)}}, \nonumber \\
\hat{b}^{\dagger} & = &  \left [\frac{d}{d\xi} (2s-1)+\frac{1}{\xi} s(2s-1)-N-\frac{1}{2} \right]\sqrt{\frac{s-1}{s(2N+1)}}, \nonumber
\end{eqnarray}
with $s=N-n$. Furthermore, the commutator between these operators,
\begin{equation}
[\hat{b}, \hat{b}^{\dagger}] = \frac{2N-2\hat{n}}{2N+1},
\end{equation}
is the same commutation relation we found for the deformed ones in (\ref{eq:crdef}), providing the number operator is defined as $\hat{n} \psi_{N,n}(\xi) = n \psi_{N,n}(\xi)$. 

This equivalence between the standard factorization and the deformed operator methods has also been highlighted in Ref. \cite{ancheyta} for systems having an infinite discrete spectrum, such as the trigonometric P\"oschl-Teller potential and the pseudoharmonic  oscillator, where it was shown that starting from any one of the algebraic methods mentioned above, one is able to construct coherent states with identical structures. Indeed, such results reinforce the meaning of ascribing to the deformed operators as those actual dynamical variables associated with the system under consideration and, particularly, the meaning of the deformed Hamiltonian (\ref{eq:spec2}) as that of a Morse-like oscillator, provided we confine ourselves to the first $N$ bound states $|0\rangle \ldots |N-1 \rangle$ of the Morse potential. \\

In similarity to harmonic oscillator coherent states (also known as field coherent states), we propose to construct our coherent states by application of a unitary displacement operator of the form 
\begin{equation}
\hat{D}_{f}(\alpha) =  \exp(\alpha \hat{A}^{\dagger}-\alpha^{\ast }\hat{A})
\label{eq:disop}
\end{equation}
to the fundamental state of the system, with $\alpha$ being a complex parameter. The above operator can be thought of as a generalization of the usual displacement operator $\hat{D}(\alpha) = \exp(\alpha \hat{a}^{\dagger}-\alpha^{\ast }\hat{a})$, where the harmonic ladder operators $\hat{a}^{\dagger}$ and $\hat{a}$ have been replaced by the deformed ones defined in (\ref{eq:defops}). Based on this idea, generalized coherent states for the Morse potential were also constructed in Ref. \cite{recamier2} by means of an approximate version of the deformed displacement operator in which the number operator is replaced by an average value, i.e, $\hat{n} \to \langle \alpha |\hat{n}|\alpha \rangle \equiv \bar{n}$. Hence the disentanglement problem of the exponential ($\ref{eq:disop}$) was circumvented by applying standard mathematical techniques, since in such a case the commutator between $\hat{A}$ and $\hat{A}^{\dagger}$ is a function of $\bar{n}$, which is obviously no longer an operator. However, the displacement operator thus obtained turned out to be approximately unitary depending on the value of the parameter $\alpha$. In this work we proceed in a different way by considering the fact that the f-deformed algebra associated with the Morse potential lends itself to a Lie algebraic treatment. \\

When referring to (\ref{eq:crdef}), it is clear that the set of operators $\{\hat{A}, \hat{A}^{\dagger}, \hat{n}, 1 \}$ forms a Lie algebra, thereby the deformed displacement operator can be disentangled as follows \cite{puri,gilmore}
\begin{eqnarray}
\hat{D}_{f}(\alpha) &= & \exp(\alpha \hat{A}^{\dagger}-\alpha^{\ast }\hat{A}) \nonumber \\
& = & \exp \left(\zeta \frac{\hat{A}^{\dagger}}{\sqrt{\chi_{a}}} \right) \left (\frac{1}{1+|\zeta|^{2}} \right)^{g(\chi_{a}\hat{n})/2\chi_{a}}\exp \left(-\zeta^{\ast} \frac{\hat{A}}{\sqrt{\chi_{a}}} \right),
\end{eqnarray}
where, for a given value of $\alpha =|\alpha|e^{i\varphi}$, we have introduced the new complex parameter $\zeta = e^{i\varphi} \tan (|\alpha|\sqrt{\chi_{a}})$, and $g(\chi_{a}\hat{n})= [\hat{A},\hat{A}^{\dagger}]$. \\

Therefore, we define our Morse-like nonlinear coherent state by applying the above deformed displacement operator upon the oscillator ground state, which gives

\begin{eqnarray}
|\zeta \rangle \equiv \hat{D}(\zeta(\alpha)) |0\rangle & = & \left(\frac{1}{1+|\zeta|^{2}} \right)^{\frac{(1-\chi_{a})}{2\chi_{a}}} \exp\left (\zeta \frac{\hat{A}^{\dagger}}{\sqrt{\chi_{a}}} \right) |0\rangle,  \label{eq:defdesp} \\ 
& = & \left(\frac{1}{1+|\zeta|^{2}} \right)^{\frac{(1-\chi_{a})}{2\chi_{a}}} \left(\sum_{n=0}^{N-1}\frac{\zeta^{n}}{\sqrt{\chi_{a}^{n}}} \frac{\hat{A}^{\dagger n}}{n!}+\sum_{n=N}^{\infty}\frac{\zeta^{n}}{\sqrt{\chi_{a}^{n}}} \frac{\hat{A}^{\dagger n}}{n!}\right)|0\rangle \nonumber \\
& \approx & \left(\frac{1}{1+|\zeta|^{2}} \right)^{\frac{(1-\chi_{a})}{2\chi_{a}}} \sum_{n=0}^{N-1}\frac{\zeta^{n}}{\sqrt{\chi_{a}^{n}}} \frac{f(n)!}{\sqrt{n!}}|n\rangle.
\label{eq:desp}
\end{eqnarray}
It is important to note the last approximation above comes from the fact that we are taking into account just the  $N$ bound states of the potential. Then, on substitution of the explicit form of $f(n)!$, namely,
\begin{equation}
f(n)! = f(1)f(2) \cdots f(n) = \sqrt{\frac{(2N)!}{(2N+1)^{n}(2N-n)!}},
\end{equation}
into Eq. (\ref{eq:desp}) we finally obtain the nonlinear coherent state
\begin{equation}
|\zeta \rangle = \sum_{n=0}^{N-1} {2N \choose n}^{1/2} \frac{\zeta^{n}}{(1+|\zeta|^{2})^{N}}|n\rangle ,
\label{eq:dcs}
\end{equation}
where ${2N \choose n} = \frac{(2N)!}{n! (2N-n)!}$. It is important to stress that since the summation is finite, these states cannot be considered as a complete set of states; in this sense they are only  approximately coherent. This issue is more evident insofar as the absolute value of the $\zeta$ paramater in (\ref{eq:dcs}) increases for a fixed number of bound states, since in such a case the influence of the continuum becomes more significant and, under this circumstance, where a complete basis would be essential, the dissociation  of the molecule may take place. Hence, considering only the low-lying region of the spectrum explains why our states are only approximate. On the other hand, one can also see in the limit $N \to \infty$ (or equivalently in the limit of $\chi_{a} $ going to zero) they contract to Glauber's coherent states $|\alpha \rangle = e^{-|\alpha|^{2}/2}\sum_{n}^{\infty}\frac{\alpha^{n}}{\sqrt{n!}}|n\rangle$. From now on, we will call the states given in (\ref{eq:dcs}) the {\it deformed Displacement Operator Coherent States} (DOCSs).

The coherent states we have obtained for the particular case of a Morse potential given by (\ref{eq:dcs}) have an algebraic structure that resembles to that of the Binomial States (BS) introduced by Stoler \etal \cite{stoler}. However, the BS are defined by the condition that the probability density is binomial. In our case the specific form of the state depends on the deformation function used to define the deformed Hamiltonian and the probability density from these states is, in general, not binomial. In this sense, our coherent states have no relationship with the Binomial ones. 

On the other hand, due to the close correspondence between the commutation relations for the deformed operators (\ref{eq:crdef}) and those for the generators of the SU(2) group \cite{carvajal}, from a purely algebraic point of view the DOCSs defined here are very similar to the so-called {\it SU(2)} or {\it spin coherent states} \cite{radcliffe}; the main difference between them being, without mentioning they belong to different contexts, the lack of completeness of our states. \\

Formerly, using the f-deformed algebra, in Ref. \cite{reca1} Morse-like coherent states were proposed as approximate eigenstates of the deformed annihilation operator. The nonclassical properties of even and odd combinations of these states  (i.e., Schr\"odinger-cat-like states) were also analyzed. For these states, a more detailed discussion about the temporal evolution of their dispersion relationships, with the help of a convenient algebraic representation of the coordinate and momentum variables in terms of the f-deformed operators, can be found in Ref. \cite{reca2}. For the same system, it is also important to mention the generalized coherent states \`a la Klauder \cite{klauder} introduced by Angelova and Hussin \cite{angelova}, which turned out to be, as well,  approximate coherent states inside the confining region of the potential. Using the Gazeau-Klauder formalism \cite{gazeau}, in Ref. \cite{roy,popov1} coherent states corresponding to the Morse potential and some of their properties were also examined. What is more, it was proposed in \cite{roknizadeh} that both the Klauder and, in some special cases, the Gazeau-Klauder coherent states can be constructed within the f-deformed oscillator formalism by choosing an appropriate deformation function (all of them described as a superposition of energy eigenstates of the corresponding quantum system) and, therefore, classified as nonlinear coherent states (more specifically, as deformed annihilation operator eigenstates). A more sophisticated approach was undertaken in Ref. \cite{benedict} from the viewpoint of supersymmetric quantum mechanics (SUSY QM) so as to describe the Morse potential and obtain its coherent states, the last ones being expressed in terms of a special basis known as the basis of the pseudo-number-states, for which the Hamiltonian is tridiagonal. However, in this regard we consider that it is much easier to carry out the factorization (\ref{eq:model1}) within the f-deformed oscillator formalism. \\

More recently, in Ref. \cite{recamier3} the displacement operator method allowed us to construct explicit expressions of nonlinear coherent states for two different systems, namely, the modified and the trigonometric P\"oschl-Teller potentials; the former supporting a finite number of bound states and the latter supporting an infinite number of bound states. In addition, a generalization in terms of eigenstates of the deformed annihilation operator was also proposed for each potential. The coherent states thus obtained displayed nonclassical features such as squeezing and, depending on the type of generalization employed, sub-Poissonian statistics.

\subsection{Deformed photon-added coherent states}

Another interesting class of nonclassical states is known as the set of {\it photon-added} states, i.e., those states of the form
\begin{equation}
|\phi,m \rangle = N_{m} \hat{a}^{\dagger m} |\phi\rangle,
\end{equation}
where $|\phi \rangle$ is an arbitrary state, $\hat{a}^{\dagger}$ is the standard creation operator, and $m$ is taken to be a non-negative integer (the number of added quanta). When the initial state $|\phi \rangle$ is a harmonic oscillator coherente state, we obtain the (PACS) introduced by Agarwal and Tara \cite{agarwal}. Much work has been devoted to the study of these states, both in their nonclassical properties (see for instance \cite{dodonov} and the references given there) and in putting forward possible schemes for their experimental realization \cite{sivakumar,zalamidas,zavatta}. Besides, some interest has been also focussed on the study of their possible generalizations by considering systems other than the harmonic oscillator \cite{safaeian,daoud,popov2}. \\

Photon added coherent states (PACS) can be explicitly written as follows:
\begin{equation}
|\alpha,m\rangle = \frac{e^{-|\alpha|^{2}/2}}{[L_{m}(-|\alpha|^{2})m!]^{1/2}}\sum_{n=0}^{\infty} \frac{\alpha^{n}\sqrt{(n+m)!}}{n!}|n+m\rangle,
\end{equation}
where $L_{m}(x)$ stands for the Laguerre polynomial of {\it m}th-order. It is known that these states display nonclassical properties such as squeezing and sub-Poissonian statistics.

In Ref. \cite{safaeian}, within the f-deformed oscillator formalism, the so-called {\it deformed photon added nonlinear coherent states} (DPANCSs) were introduced by using the definition 
\begin{equation}
|\alpha,f,m\rangle = N_{\alpha}^{m}\hat{A}^{\dagger m} |\alpha,f\rangle,
\end{equation}
where, unlike the harmonic case, these states are constructed by applying iteratively the deformed creation operator $\hat{A}^{\dagger}$ upon an initial state that is considered to be an eigenstate of the deformed annihilation operator $\hat{A}$.

Inspired by the above-mentioned studies we propose another way to generalize the photon added coherent states, and that is to build them by the iterative application of the deformed creation operator upon the DOCS given by Eq. (\ref{eq:dcs}), that is, 
\begin{eqnarray} 
|\zeta,m\rangle & = & C_{\zeta,m} \hat{A}^{\dagger m} \hat{D}(\zeta) |0\rangle \nonumber \\
& = & C_{\zeta,m} \hat{A}^{\dagger m}|\zeta \rangle.
\end{eqnarray}
More explicitly, our {\it deformed photon added coherent states} (DPACSs) for the Morse potential take the following form
\begin{equation}
|\zeta,m\rangle = \frac{C_{\zeta,m}}{(1+|\zeta|^{2})^{N}}\sum_{n=0}^{N-1-m} {2N \choose n+m}^{1/2} \frac{(n+m)!}{n!}\zeta^{n} |n+m\rangle,
\label{eq:dpacs}
\end{equation}
where the normalization factor is given by
\begin{equation}
C_{\zeta,m} = (1+|\zeta|^{2})^{N}\left( \sum_{n=0}^{N-1-m}{2N \choose n+m} \left(\frac{(n+m)!}{n!}\right)^{2} |\zeta|^{2n} \right)^{-1/2}.
\end{equation}
Again, since we are dealing with a potential having a finite number of bound states, the summation in (\ref{eq:dpacs}) must end at $n= N-1-m$.

\section{The Wigner function of Morse-like coherent states}

We are now in a position to proceed to the calculation of the Wigner function of the states $|\zeta \rangle$ and $|\zeta,m\rangle$. First of all, we start by recalling that if a system is in the state $\phi(x)$, the Wigner distribution function, in terms of position-space wavefunctions, is defined as \cite{kim} 
\begin{equation}
W(x,p,t) = \frac{1}{2\pi \hbar} \int_{-\infty}^{\infty} e^{-ipy/\hbar} \phi^{\ast}(x-y/2,t) \phi(x+y/2,t)dy,
\label{eq:fwigner}
\end{equation}
where $x$ and $p$ are simply variables of position and momentum, respectively, and have no operator properties.

In virtue of the set of relations (\ref{eq:acdef}) and (\ref{eq:ladder}), let us identify the number states $\{|n\rangle \}$ in (\ref{eq:dcs}) with the Morse eigenfunctions $\psi_{N,n}(x)$ given in (\ref{eq:wavefuncs}), in order that the wave function corresponding to the nonlinear coherent states $|\zeta \rangle$ might be 
\begin{equation}
\Psi_{\zeta} (x)  = \sum_{n=0}^{N-1} {2N \choose n}^{1/2} \frac{\zeta^{n}}{(1+|\zeta|^{2})^{N}}\psi_{N,n} (x).
\label{eq:wavecs}
\end{equation}
This identification is essential if one wants to give an account of the influence of potential's asymmetric profile in phase space (as it will be seen below in section of numerical results).

So, substitution of (\ref{eq:wavecs}) in (\ref{eq:fwigner}) gives 

{\setlength\arraycolsep{2pt}
\begin{eqnarray}
W_{\zeta}(x,p) & = &  \frac{1}{2 \pi \hbar}\int_{-\infty}^{\infty} e^{-ipy/\hbar}\Psi_{\zeta}^{\ast}(x-y/2)\Psi_{\zeta}(x+y/2)dy \label{eq:calwig} \\
& = & \frac{1}{2 \pi \hbar (1+|\zeta|^{2})^{2N}}\sum_{n,k}^{N-1} {2N \choose n}^{1/2} {2N \choose k}^{1/2} \zeta^{n} \zeta^{\ast k} \times {} \nonumber \\  & & {} \times \int_{-\infty}^{\infty} e^{-ipy/\hbar}\psi_{N,k}^{\ast}(x-y/2)\psi_{N,n}(x+y/2)dy. \nonumber
\end{eqnarray}}
As in Ref. \cite{frank}, where the authors studied the Wigner function of bound eigenstates  for the Morse potential, in order to calculate the last integral in (\ref{eq:calwig}) it is convenient to use the Morse variable (\ref{eq:mvariable}) together with the change of variable $u=e^{\beta y/2}$. Thus, on substituting (\ref{eq:wavefuncs}) into (\ref{eq:calwig}), we obtain:
{\setlength\arraycolsep{2pt}
\begin{eqnarray} 
\fl W_{\zeta}(x,p) = \frac{1}{\pi \hbar \beta (1+|\zeta|^{2})^{2N}} \sum_{n,k}^{N-1} {2N \choose n}^{1/2} {2N \choose k}^{1/2} \zeta^{n} \zeta^{\ast k} C_{N,n}C_{N,k} \xi^{2N-n-k} {}  \label{eq:calwin2}\\ 
  {} \times \int_{0}^{\infty} \exp \left[-\frac{\xi}{2}\left (u+\frac{1}{u} \right) \right] L_{k}^{2N-2k}(\xi u)L_{n}^{2N-2n}(\xi u^{-1})u^{n-k-1-2ip/\hbar \beta}du. \nonumber
\end{eqnarray}}
Then, if we express the associated Laguerre polynomials in terms of their finite series
\begin{equation}
L_{n}^{k}(x)= \sum_{m=0}^{n}(-1)^{m} {n+k \choose n-m}\frac{x^{m}}{m!},
\end{equation}
it is possible to integrate Eq.(\ref{eq:calwin2}), using Eq. (3.471.9) in Ref. \cite{grads}, that is 
\begin{equation*}
\int_{0}^{\infty}x^{\nu-1}\exp \left(-\frac{\beta}{x}-\gamma x \right) dx = 2\left(\frac{\beta}{\gamma} \right)^{\nu/2}K_{\nu}(2\sqrt{\beta \gamma}),
\end{equation*}
to get the final result 
{\setlength\arraycolsep{2pt}
\begin{eqnarray}\label{eq:wigdcs} 
W_{\zeta}(x,p) & = & \frac{2}{\pi \beta \hbar (1+|\zeta|^{2})^{2N}} \sum_{n,k=0}^{N-1}{2N \choose n}^{1/2}{2N \choose k}^{1/2}\zeta^{\ast k}\zeta^{n} \\ & & {}  
\times C_{N,n}C_{N,k} \xi^{2N-n-k}  \sum_{r=0}^{k}\sum_{s=0}^{n}{2N-k \choose k-r}{2N-n \choose n-s}  \nonumber \\ & & {}  \times \frac{(-\xi(x))^{r+s}}{r!s!}K_{r+n-(s+k)-2ip/\hbar \beta}(\xi(x)),  \nonumber
\end{eqnarray}}
where $K_{\nu}(\xi)$ are the modified Bessel functions of the third kind. We have thus gotten the corresponding Wigner function of the DOCSs for a potential having $N-1$ bound states. In a similar manner it is possible to determine the distribution function corresponding to the DPACSs defined in Eq. (\ref{eq:dpacs}). The result is:

{\setlength\arraycolsep{2pt}
\begin{eqnarray}\label{eq:wigadded}
\fl W_{\zeta,m}(x,p)  =  \frac{2 C_{\zeta,m}^{2}}{\pi \beta \hbar (1+|\zeta|^{2})^{2N}} \sum_{n,k=0}^{N-1-m}{2N \choose n+m}^{1/2}{2N \choose k+m}^{1/2}\frac{(n+m)!}{n!}\frac{(k+m)!}{k!} \zeta^{\ast k}\zeta^{n} \nonumber \\   {}  \times C_{N,n+m}C_{N,k+m} \xi^{2N-2m -n-k}  \sum_{r=0}^{k+m}\sum_{s=0}^{n+m}{2N-k-m \choose k+m-r}{2N-n-m \choose n+m-s}\nonumber \\  {}  \times \frac{(-\xi(x))^{r+s}}{r!s!} K_{r+n-(s+k)-2ip/\hbar \beta}(\xi(x)). 
\end{eqnarray}}
In the case of $\zeta=0$ the last expression is nothing but the Wigner distribution function for the bound eigenstates of the Morse potential calculated by Frank \etal \cite{frank}, that is, 
{\setlength\arraycolsep{2pt}
\begin{eqnarray}
W_{0,m}(x,p) & = & \frac{4 m! (N-m)}{\pi \hbar \Gamma(2N-m+1)}  \xi^{2N-2m}  \sum_{r=0}^{m}\sum_{s=0}^{m}{2N-m \choose m-r}{2N-m \choose m-s}\nonumber \\ & & {}  \times \frac{(-\xi(x))^{r+s}}{r!s!} K_{r-s-2ip/\hbar \beta}(\xi(x)).
\end{eqnarray}}
In particular, for $m=0$, it is easy to see from this expression that the Wigner function of the ground state takes the form 
\begin{equation}
W_{0,0}(x,p)=\frac{2}{\pi \hbar \Gamma(2N)}\xi^{2N}K_{-2ip/\beta \hbar}(\xi(x)).
\end{equation}
This result is also found in Ref. \cite{frank} as expected. \\

In what follows we shall employ the results so far obtained in order to compute and examine the behavior of the Wigner functions of our Morse-like coherent states. From now on we set $\hbar =\Omega= \mu =1$ for simplicity.  

\section{Numerical results}

\begin{figure}[h!]
\begin{center}
\includegraphics[width=11cm, height=5cm]{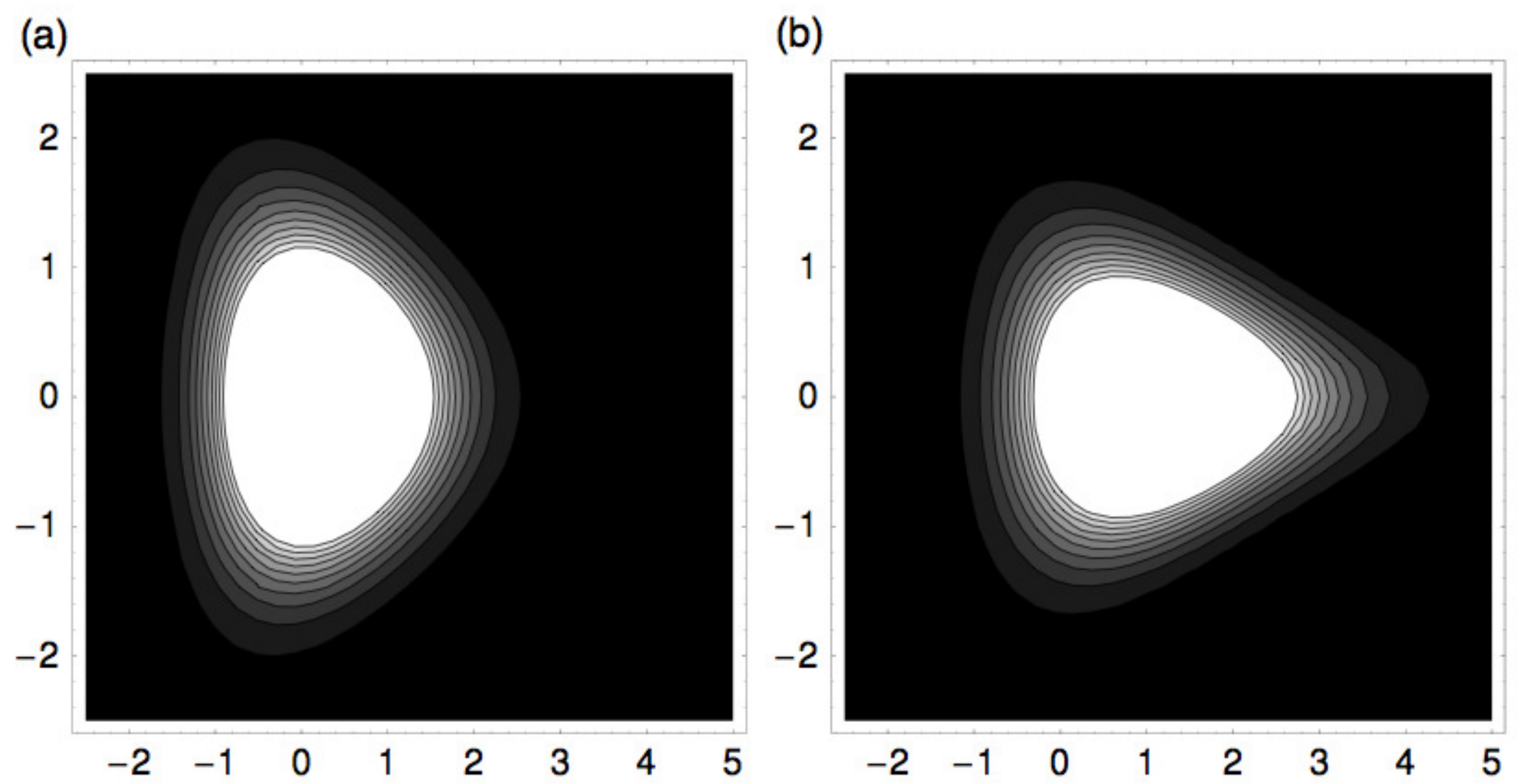} 
\caption{Contour plots of the Wigner functions of (a) the ground state $|\zeta =0 \rangle$, $\langle \hat{n} \rangle =0$, and (b) the coherent state $|\zeta \rangle$, $\langle \hat{n} \rangle= 0.25$, corresponding to a Morse potential with $N=10$ bound states. The abscissa is the position $x$ and the ordinate is the momentum $p$. Bright color corresponds to large values of $W(x,p)$.}
\label{fig:ccmor}
\end{center}
\end{figure}

The contour plot of the Wigner functions corresponding to the DOCSs from Eq. (\ref{eq:wigdcs}) for $\langle \hat{n} \rangle =0$ and $\langle \hat{n} \rangle =0.25$ are shown in Figs. \ref{fig:ccmor} (a) and (b), respectively, for a Morse potential having $N=10$ ($\chi_{a} \approx 0.05$) bound states. The phase space distribution in (a) is no longer gaussian in form, unlike the harmonic oscillator case where the contour plot of the Wigner function shows concentric circles; it appears slightly squeezed in the $x$ direction. In (b) we see that the Wigner function for the DOCS with $\langle \hat{n} \rangle = 0.25$  appears to be slightly squeezed in the $p$ direction and much more elongated in the $x$ direction. \\

\begin{figure}[h!]
\begin{center}
\includegraphics[width=9cm, height=9cm]{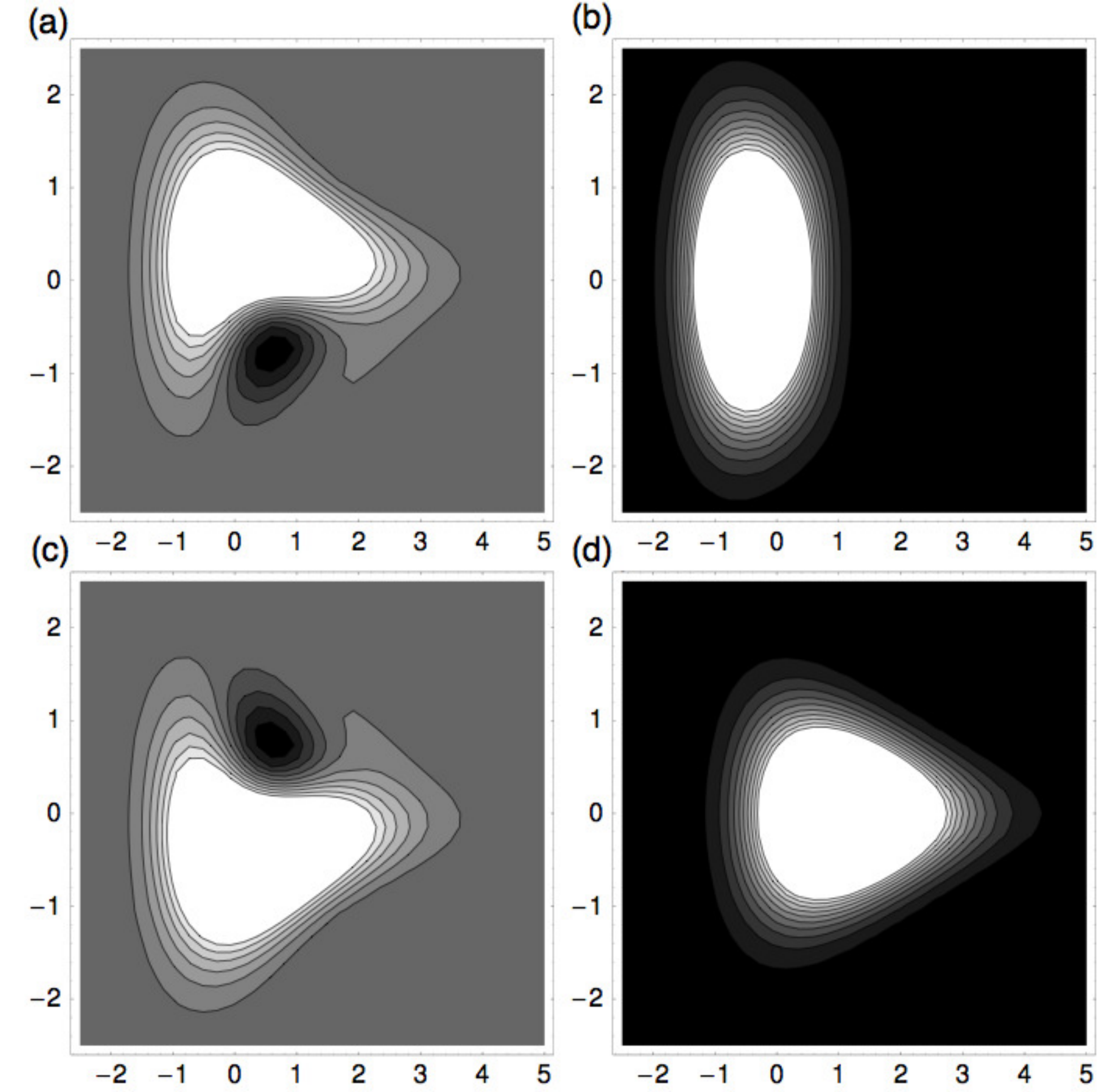} 
\caption{Temporal evolution of the Wigner function of the coherent state $|\zeta(t)\rangle = \hat{U}_{D}(t) |\zeta(0)\rangle$, for $\langle \hat{n} \rangle= 0.25$, corresponding to a Morse potential with $N=10$ bound states. Frames (a)-(d) correspond to time instants $t=\tau/4$, $\tau/2$, $3\tau/4$ and $ \tau$, respectively. The abscissa is the position $x$ and the ordinate is the momentum $p$.}
\label{fig:cctmor}
\end{center}
\end{figure}

The temporal evolution of the Wigner function is shown in Fig. (\ref{fig:cctmor}).
By application of the time evolution operator $\hat{U}(t) = e^{-i\hat{H}_{f}t/\hbar}$ upon the coherent state at $t=0$  using the deformed Hamiltonian given by (\ref{eq:spec2}) we get the coherent state at time $t$, 
\begin{equation}
|\zeta (t)\rangle = \hat{U}(t)|\zeta(0) \rangle  = e^{-i \Omega t(\hat{n}+1/2-(\hat{n}+1/2)^{2}\chi_{a}-\chi_{a}/4)}|\zeta (0) \rangle,  
\label{docsevol}
\end{equation}
where the initial state $|\zeta(0) \rangle$ is given by (\ref{eq:dcs}). We calculated the phase space distribution for different times $t=\tau/4$, $\tau/2$, $3\tau/4$ and $\tau$; $\tau= 2\pi/\Omega \chi_{a}$ being an estimated period for which the phase-space distribution approaches that in Fig. \ref{fig:ccmor} (b). Then, as time elapses, say $t=\tau/4$, we see that due to the quadratic term appearing in the time evolution operator the Wigner function acquires negative values (see dark area in Fig. \ref{fig:cctmor} (a)) which means that the nonlinear coherent state presents a nonclassical conduct. At $t=\tau/2$ (see Fig. \ref{fig:cctmor} (b)), on account of the potential's barrier for $x < 0$, it is observed that the distribution appears highly squeezed in the $x$ direction and consequently more elongated in the $p$ direction. The Wigner function at $t=3\tau /4$ (see Fig. \ref{fig:cctmor} (c)) represents the reflected image, with respect the $x$ axis, of that in Fig. \ref{fig:cctmor} (a). Finally, our coherent state regains its initial distribution after a time interval of $t \approx \tau$. Such a behavior repeats itself periodically. \\

\begin{figure}[h!]
\begin{center}
\includegraphics[width=11cm, height=13cm]{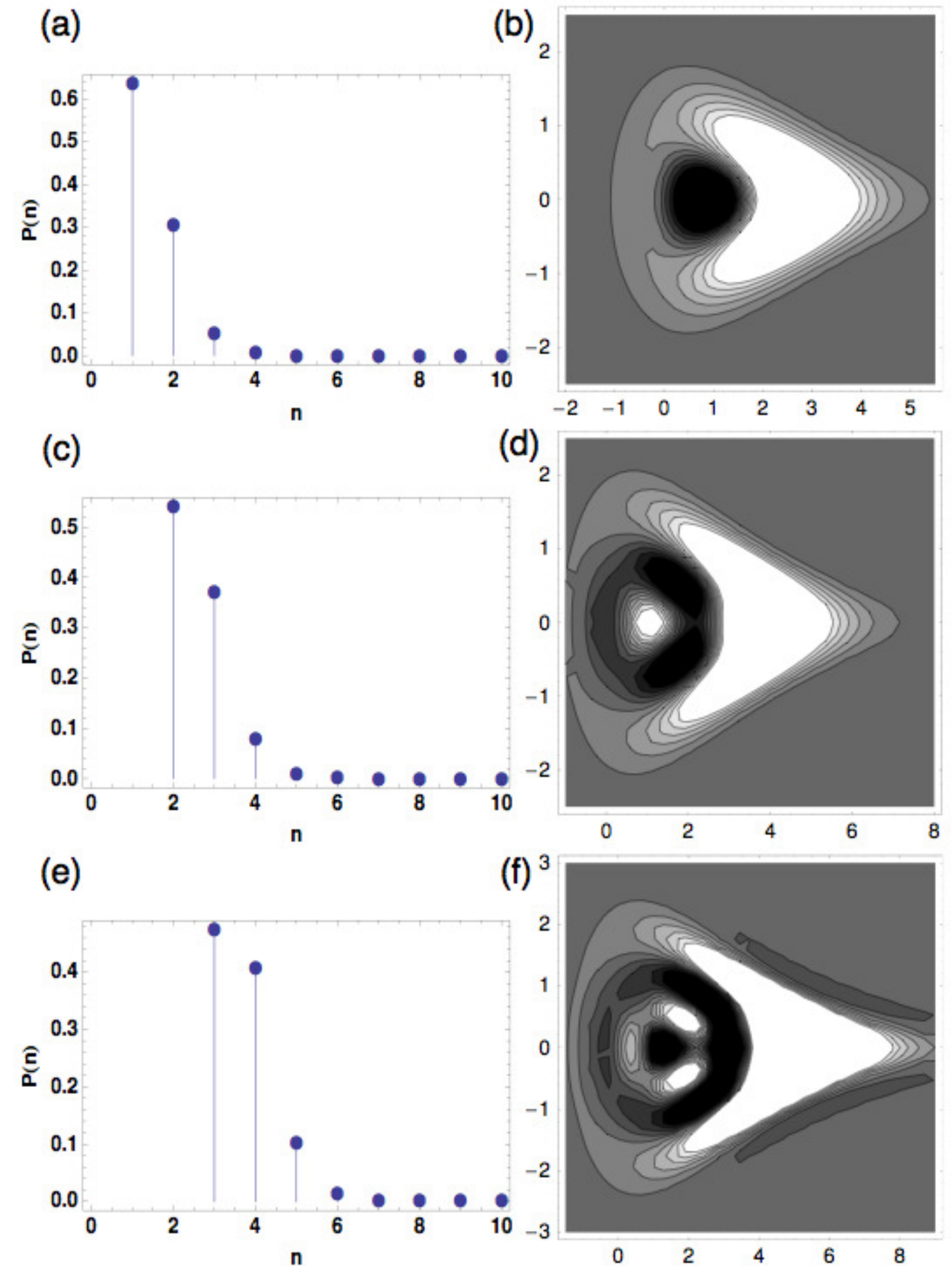} 
\caption{Occupation number distributions (frames (a), (c) and (e)) and the corresponding contour plots of the Wigner function (frames (b), (d) and (f))  of deformed photon added coherent states (DPACSs) for $m=1,2$ and $3$. The dimensionless coordinate $x$ and momentum $p$ are depicted by the abscissa and the ordinate, respectively. Dark areas correspond to negative values of $W(x,p)$ and bright areas correspond to positive values.}
\label{fig:ccadded}
\end{center}
\end{figure}

Next, we show in Fig. \ref{fig:ccadded} the phase-space distribution of the DPACSs with the help of the Wigner function (\ref{eq:wigadded}). In the same figure it is also shown the occupation number distribution, $P(n)=|\langle n |\zeta,m\rangle|^{2}$, as a function of $n$ for different values of $m$, where one can see those anharmonic excitations that play an important role in the construction and/or evolution of our deformed photon-added coherent states. Once again, for the sake of comparison, we have taken into account a potential with $N=10$ bound states and the $\zeta$ parameter is taken to be a constant whose value corresponds to an average $\langle \hat{n} \rangle =0.25$.

It is clear that in the expansion of the DPACSs $|\zeta,m\rangle$ in terms of number states $|n\rangle$ the states $|0\rangle$, $|1\rangle$,\ldots, $|m-1\rangle$ do not contribute to the summation. This is illustrated in Figs. \ref{fig:ccadded} (a), (c) and (e) for $m=1$, $2$ and $3$, respectively. For the case of $m=1$, i.e., in the absence of the ground state, the Wigner function displays a distribution whose form (see Fig. \ref{fig:ccadded} (b)) is somewhat different to the one exhibited in Fig. \ref{fig:ccmor} (b). The former, in comparison with the latter, takes negative values being represented by the dark area in Fig. \ref{fig:ccadded} (b).  Clearly, this is already a signature of the nonclassical nature of our photon-added coherent states. The loss of classicality of these states seems to be more evident as $m$ increases, say $m=2$ and $3$, as one can see from Figs. \ref{fig:ccadded} (d) and (f). One can also see that not only does the Wigner function take negative values, but it also appears to be more elongated towards the $x$ direction. This implies that with the increase of $m$, the uncertainty of the associated position-space wavefunction becomes larger. From Figs. \ref{fig:ccadded} (a), (c) and (e), note that we have concentrated our attention on those energy regions where the contribution of the continuum can be neglected.\\

\begin{figure}[h!]
\begin{center}
\includegraphics[width=14cm, height=9cm]{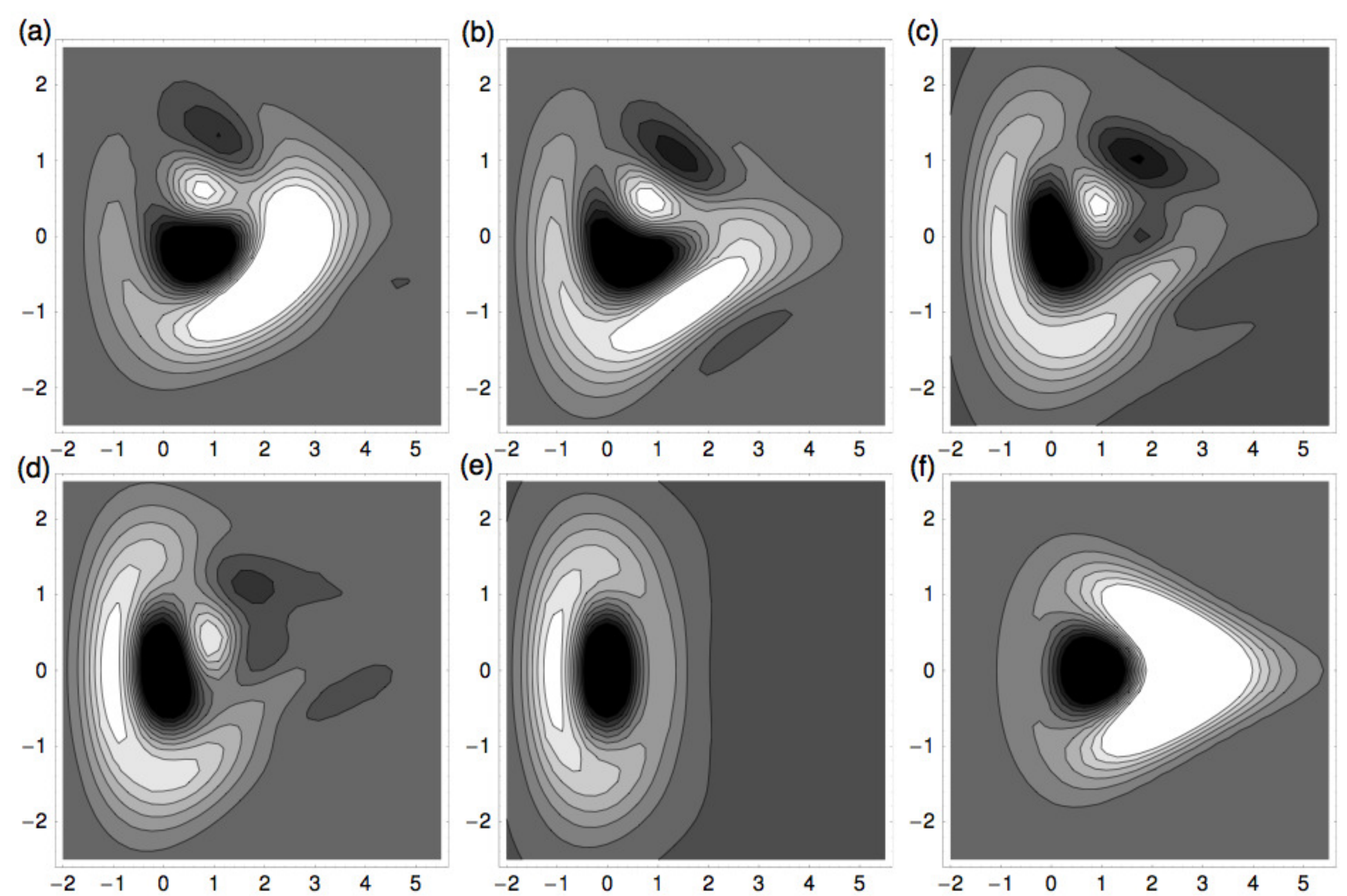} 
\caption{Evolution of the Wigner function of a deformed photon added coherent state with $m=1$. Frames (a) to (f) correspond to time instants $t=\tau/8$, $\tau/4$, $3\tau/8$, $7\tau/16$, $\tau/2$ and $\tau$, respectively. Dark areas correspond to negative values of $W(x,p)$ and bright areas correspond to positive values.}
\label{fig:ccaddedt}
\end{center}
\end{figure}

Let us finally examine the dynamics of the Wigner function for the DPACSs that is displayed in Fig. \ref{fig:ccaddedt}. For this task, when referring to Eq. (\ref{docsevol}), we have chosen as initial state $|\zeta (0) \rangle$ the one defined by Eq. (\ref{eq:dpacs}) provided that $m=1$, that is, the same state as that represented in Fig. \ref{fig:ccadded} (b). So then, the frames \ref{fig:ccaddedt}(a) to (f) correspond to contour plots of the Wigner distribution function at $t=\tau /8$, $\tau/4$, $3\tau/8$, $7\tau/16$, $\tau/2$ and $\tau$, respectively. It is evident that, as time elapses, the results of the photon-added case are not the same as those of the deformed displacement operator coherent states. We found that the quadratic term of the time evolution operator, in conjunction with the photon-added effect, leads to a fast spreading of state's phase space distribution.  A couple of deformed hills are barely discerned for the first time instants $t=\tau /8$, $\tau/4$, $3\tau/8$ and $7\tau/16$, one of those being much smaller than the other one (see bright area in Figs. \ref{fig:ccaddedt}(a)-(d)). At $t =\tau/2$ (Fig. \ref{fig:ccaddedt} (e)), due to potential's barrier effect, the state is not quite refocused into a single component like that at $t=0$. The initial state is completely regained after a time $t=\tau$ (see Fig. \ref{fig:ccaddedt} (f)).

\section{Conclusions}

In this work we have analyzed coherent states for the Morse potential (Morse-like coherent states) and examined their behavior in the phase space from the viewpoint of the Wigner distribution function. More specifically, using the f-deformed oscillator formalism, such states were constructed considering two types of generalization, as deformed displacement operator coherent states (DOCSs) and as deformed photon added coherent states (DPACSs). For the states thus constructed, analytical expressions of their Wigner functions were also obtained. The analysis based upon the above-mentioned distribution function revealed the following results: (i) After applying the deformed displacement operator on the ground state of the system, the resulting coherent states (DOCSs) can be considered as a well-localized states on phase space; however, they will unavoidable be influenced by the nonlinearity of the Hamiltonian at certain stages of their evolution, whereby such states turn out to be, in general, nonclassical states. (ii) The DPACSs  exhibit a nonclassical behavior all the time. Both the photon-added effect and the quadratic term in the time evolution operator lead to a noticeable spreading of the states' phase space distribution. Indeed, it is found that the larger the number of added quanta $m$ is, the more evident the loss of classicality of these states is.

\ack We acknowledge partial support from DGAPA UNAM through project IN120909.

\end{document}